\documentclass[aps,prl,twocolumn,groupedaddress,showpacs]{revtex4}
\usepackage{graphicx}
\usepackage{amsmath}

\begin{document}

\title{Charge puddles in a completely compensated topological insulator}

\author{C. W. Rischau$^1$, A. Ubaldini$^2$, E. Giannini$^2$, and C. J. van der Beek$^1$\footnote{Corresponding author: kees.vanderbeek@polytechnique.edu}}
\affiliation{$^1$Laboratoire des Solides Irradi\'es, Ecole Polytechnique, CNRS, CEA, Universit\'e Paris-Saclay, 91128 Palaiseau Cedex, France\\
$^2$University of Geneva, CH-1211 Geneva 4, Switzerland\\
}



\date{\today}

\begin{abstract}
Compensation of intrinsic charges is widely used to reduce the bulk conductivity of 3D topological insulators (TIs). Here we use low temperature electron irradiation-induced defects paired with \textit{in-situ} electrical transport measurements to fine-tune the degree of compensation in Bi$_2$Te$_3$. The coexistence of electrons and holes at the point of optimal compensation can only be explained by bulk carriers forming charge puddles. These need to be considered to understand the electric transport in compensated TI samples, irrespective of the method of compensation.
\end{abstract}


\maketitle
Since the discovery of three-dimensional topological insulators (TIs), electrical transport studies performed to put their topologically protected surface states into evidence have been plagued by the high bulk conductivity of these materials \cite{Qu:2010,Hor:2010,Butch:2010,Eto:2010,Ren:2010,Xiong:2012,Ren:2012}. While TI materials such as Bi$_2$Te$_3$, Bi$_2$Se$_3$ or Sb$_2$Te$_3$ have bulk bandgaps of typically a few 100 meV and should therefore behave as band insulators, native defects always dope these materials thereby shifting the chemical potential into the bulk conduction or valence band \cite{Scanlon:2012}. Most bulk TI samples can thus be considered as heavily doped semiconductors and consequently display an undesired metallic conduction. Despite tremendous efforts, even the cleanest TI bulk samples still have carrier densities as high as $10^{16}-10^{17}$ cm$^{-3}$. Since it does not seem realistic that insulating Bi-based bulk TI samples can be achieved by reducing only the defect density \cite{Brahlek:2014B}, it is necessary to compensate the carriers already present to push the Fermi level inside the bandgap. In Bi$_2$Te$_3$, compensation can be realized either by chemical doping, e.g., Bi$_2$Te$_{3-x}$Se$_x$ \cite{Ren:2010,Xiong:2012,Ren:2012,Akrap:2014}, or as recently proposed, by doping using irradiation-induced defects \cite{Rischau:2013}. Irradiations performed on Bi$_2$Te$_3$ at room temperature showed that the created defects act as electron donors and can thus be used to change the conduction from $p$- to $n$-type. However, the precision to adjust the degree of compensation was limited due to the elevated irradiation temperature.\\
Here we use low-temperature electron irradiation paired with \textit{in-situ} electrical transport measurements to tune the defect density and, with it, the carrier concentration in Bi$_2$Te$_3$ with high precision. This method offers a unique way to study the effects of compensation in a semiconductor as it allows the continuous change of the degree of compensation on the \textit{same} sample. We find that at the point of optimal compensation or charge neutrality point (CNP), electron- and hole-type carriers coexist. This coexistence is evidence for bulk carriers forming charge puddles, as proposed by the theory of completely compensated semiconductors \cite{Shklovskii:1972,Shklovskii:1984}. Charge puddles have been predicted to be responsible for the small bulk resistivity of TIs \cite{Skinner:2012}.\\
\begin{figure}
	\centering
		\includegraphics[width=0.48\textwidth]{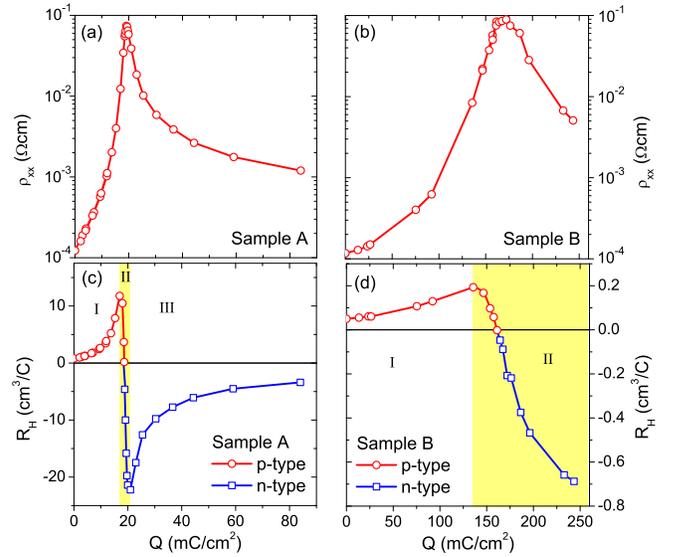}
	\caption{\textbf{(a)} \textbf{(b)} Resistivity $\rho_{xx}$ and \textbf{(c)} \textbf{(d)} low-field Hall coefficient $R_H$ of samples A and B measured \textit{in-situ} at 4 K as a function of electron dose $Q$. Labels I, II and III mark the dose regions in which the carrier densities have been calculated assuming a one-carrier-type (I and III) or a two-carrier-types model (II).}
	\label{fig:figure1}
\end{figure}
We have irradiated initially \textit{p}-type Bi$_2$Te$_3$ single crystals at low temperature (irradiation temperature 4 - 10 K) with 2.5 MeV electrons at the \textit{SIRIUS} Pelletron accelerator facility of the \textit{Laboratoire des solides irradi\'es}. After irradiation to selected electron doses $Q$, the Hall resistivity $\rho_{yx}$ and resistivity $\rho_{xx}$ were measured \textit{in-situ} at 4 K up to magnetic fields of 3 T with an AC excitation in the four-point geometry. The Bi$_{2}$Te$_3$ single crystals had lateral dimensions of $1\times3$ mm$^2$ and thicknesses between 10 and 50 $\mu$m.\\
Figs. \ref{fig:figure1} (a) and (b) depict the resistivity $\rho_{xx}$ of two samples with initial hole densities of $p_0=4.2\times10^{18}$ (sample A) and $4.4\times10^{19}$ cm$^{-3}$ (sample B), respectively, as a function of electron dose $Q$. Similar to room temperature irradiations \cite{Rischau:2013}, $\rho_{xx}$ shows a maximum, which at low temperature amounts to an increase by three orders of magnitude. The maximum values of 74 and 90 m$\Omega$cm for samples A and B, respectively, are comparable to those obtained on non-metallic Bi$_2$Te$_3$ samples (12 m$\Omega$cm) cut from crystals grown with a weak compositional gradient \cite{Qu:2010} or lightly doped Bi$_{1.9}$Tl$_{0.1}$Te$_3$ samples (28 m$\Omega$cm) \cite{Chi:2013}. However, they are lower than those obtained for heavy chemical doping in the Bi$_2$Te$_{3-x}$Se$_{x}$ system ($1-20$ $\Omega$cm for $x=0.9-1$) \cite{Ren:2010,Akrap:2014,Shekhar:2014}. It should be noted that although Bi$_2$Te$_2$Se has the same crystal structure as Bi$_2$Te$_3$ and is commonly referred to as \textit{chemically doped} Bi$_2$Te$_3$, the electronic structures of the two compounds are in fact quite different, particularly near the band edges. The band gap in Bi$_2$Te$_2$Se (0.3 eV) is twice as large than in Bi$_2$Te$_3$ (0.15 eV) and the band extrema in Bi$_2$Te$_2$Se are located at the $\Gamma$ point whereas they are located at off-symmetry points in Bi$_2$Te$_3$ \cite{Shi:2014}.\\
Figs. \ref{fig:figure1} (c) and (d) show the dose dependence of the low-field Hall coefficient $R_H$, which reveals three different regimes. At low doses (regime I), $R_H(Q)$ increases until it reaches a maximum. In regime II (area shaded in yellow), $R_H(Q)$ decreases until it eventually vanishes, at the dose at which $\rho_{xx}$ shows its maximum, and then changes its sign. For sample A a third regime is observed in which $R_H$ increases again, but remains negative.\\
\begin{figure}
	\centering
		\includegraphics[width=0.48\textwidth]{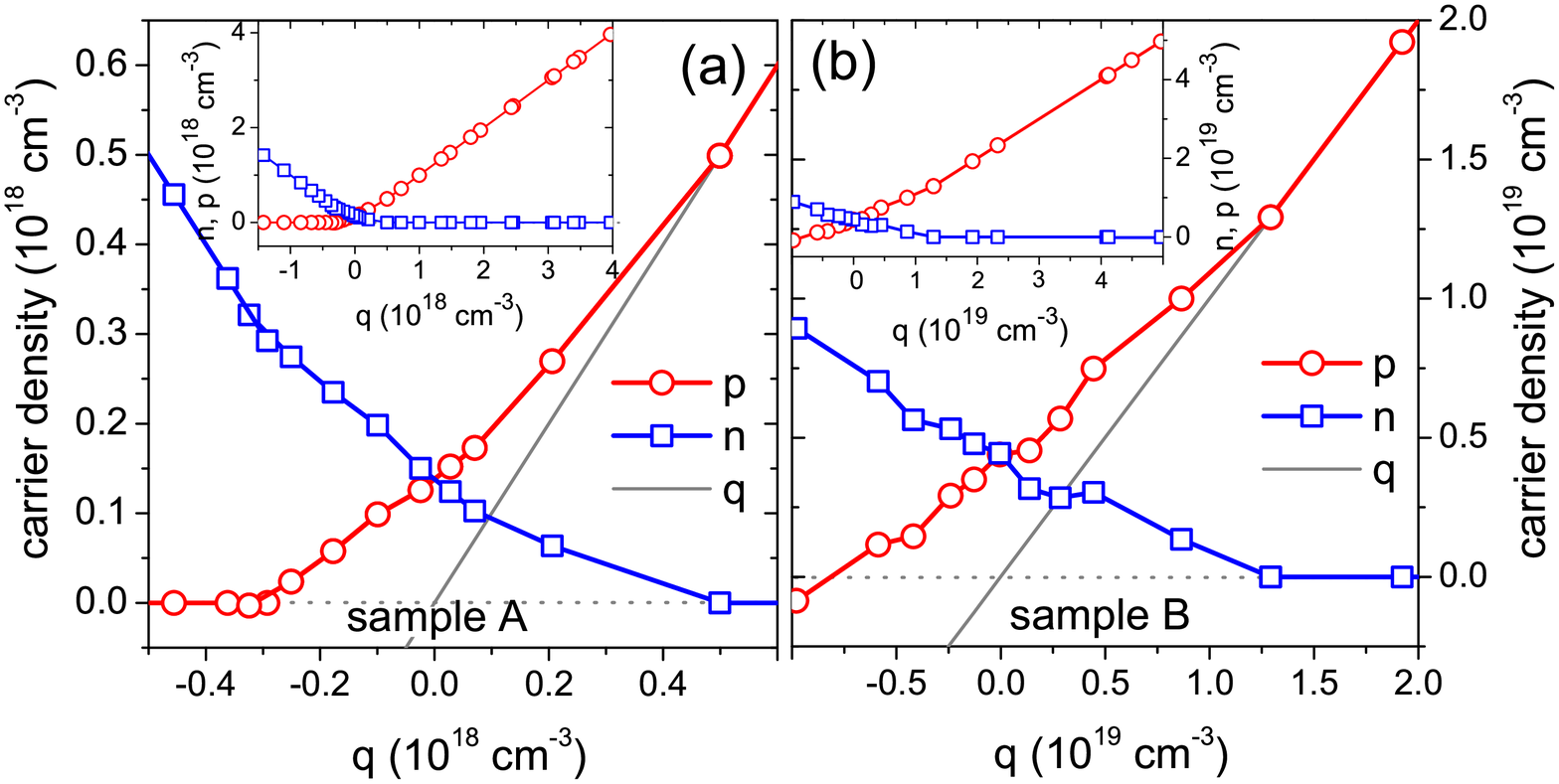}
	\caption{Hole and electron densities $p$ and $n$, respectively, calculated from the Hall coefficient $R_H$ in regime II using the model described in the text as a function of relative carrier concentration $q$ for samples \textbf{(a)} A and \textbf{(b)} B. The insets depict $p$ and $n$ in the entire $q$ range.}
	\label{fig:figure2}
\end{figure}
In conventional finite gap semiconductors, $R_H$ is determined by either holes or electrons and is inversely proportional to the carrier concentration. In regime I, $R_H$ can be described by such a single-carrier-type model of the form $R_H\propto 1/ep$ with $p$ the density of the hole type carriers. Analogously, regime III can be described by $R_H\propto 1/en$ with $n$ the density of the electrons. In regime II, $R_{H}(Q)$ vanishes around the maximum of $\rho_{xx}$. This can only be explained by a two-carrier-type model featuring the \textit{coexistence of electrons and holes}. Assuming only a single carrier type of density $\propto 1/eR_H$ would result in an infinite carrier density. Using a simplified two carrier model (see Supplemental Material, Section B for a detailed description), $R_H$ is given by 
\begin{equation}
R_H=r\frac{p-n}{e(p+n)^2}=r\frac{q}{e(p+n)^2}
\label{eq:Eq1}
\end{equation}
with the relative carrier concentration $q=p-n$ and the Hall factor $r$. At the CNP ($R_{H}=0$), the hole-type carriers initially present in the unirradiated samples are completely compensated by the donors introduced by irradiation. Although $p$ and $n$ change as a function of $Q$, the total electric charge has to be conserved during irradiation. Assuming the formation of donor-type defects during irradiation, the \textit{equation of charge neutrality} can be written as
\begin{equation}
q = p - n = p_0 - n_0 - \gamma Q
\label{eq:Eq2}
\end{equation}
with $p_0=p(Q=0)$, $n_0=n(Q=0)$ and $\gamma$ the change of the relative carrier concentration per unit dose (see Supplemental Material, Section B for more details). Combining Eqs. \ref{eq:Eq1} and \ref{eq:Eq2} allows one to calculate $p$ and $n$ which are plotted in Figs. \ref{fig:figure2} (a) and (b) as a function of $q$. Both holes and electrons are present in regime II above ($q>0$) and below ($q<0$) the CNP. The carrier densities extracted at the CNP amount to $p_{CNP}=n_{CNP}=1.4\times$10$^{17}$ and $4\times$10$^{18}$ cm$^{-3}$ for samples A and B, respectively. The observed $R_H(Q)$, in particular the smooth crossing of $R_H(Q)=0$ can thus be explained with a model based on the coexistence of electrons and holes. A similar behaviour of $R_H$ accompanied by a strong increase of $\rho_{xx}$ has been observed on gated graphene \cite{Wiedmann:2011} or TI thin films \cite{Chen:2011,Steinberg:2011,Kim:2012,Lang:2013} when the carrier density and conduction is tuned from $p$- to $n$-type by changing the gate voltage (\textit{ambipolar field effect}). In order to obtain further information on the nature of the carriers that coexist at the CNP in irradiated Bi$_2$Te$_3$, one can transform the calculated (bulk) carrier densities $n_{CNP}$ into areal carrier densities $n_{CNP}^{2D}=t \times n_{CNP}$ with $t$ the thickness of the samples ($t=30$ and 17 $\mu$m for samples A and B). This yields $n_{CNP}^{2D}=4.2 \times 10^{14}$ and $6.8 \times 10^{15}$ cm$^{-2}$ for samples A and B, respectively. One can estimate that the surface states on Bi$_2$Te$_3$ samples can only host about $3\times10^{13}$ cm$^{-2}$ carriers \footnote{The energy dispersion of massless Dirac fermions is described by $E=\pm\hbar v_{F}k$ with $v_{F}$ the Fermi velocity and $k$ the wave vector. The carrier density associated with one Dirac cone amounts to $k_{F}^{2}/4\pi$ with $k_F$ the Fermi wave vector, i.e., for two surfaces one obtains a total surface carrier density of $n_{S}=k_{F}^{2}/2\pi=\pi \epsilon_{F}^{2}/h^{2}v_{F}^{2}$. Assuming a Fermi energy lying at the bottom of the conduction band $\epsilon_F=0.34$ eV (measured from the Dirac point) \cite{Chen:2009} and a Fermi velocity of $v_F\approx 5 \times 10^{5}$ m/s \cite{Chen:2009,Qu:2010} for the surface carriers, one obtains $n_{S}\approx3\times10^{13}$ cm$^{-2}$.}, i.e., orders of magnitude lower than the value $n_{CNP}^{2D}$ found at the CNP. 
The analysis of the dose-dependence of the Hall effect and zero-field resistivity thus show that the holes present in the unirradiated Bi$_2$Te$_3$ samples can be compensated by irradiation-induced donor-type carriers. Although the bulk conductivity at the CNP has reduced by three orders of magnitude, the majority of the charge carriers coexisting at the CNP seem to be of bulk origin ($n_{CNP}^{2D} >> n_{S}$). In order to check for signatures of the surface states, we further performed \textit{in-situ} magnetoresistance measurements after each irradiation dose.\\
\begin{figure*}
	\centering
		\includegraphics[width=1.0\textwidth]{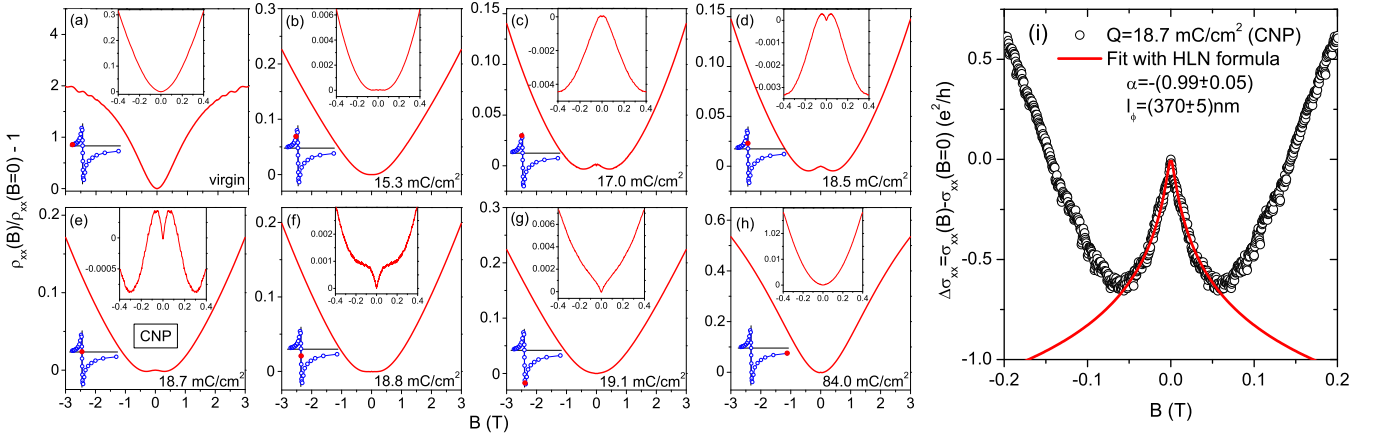}
	\caption{\textbf{(a)} - \textbf{(h)} Normalized magnetoresistance $\rho_{xx}(B)/\rho_{xx}(B=0)-1$ of sample A measured \textit{in-situ} at 4 K ($\boldsymbol{B}\parallel\boldsymbol{c}$-axis) after irradiation to different electron doses $Q$. The upper insets show a close up at low fields. The lower insets depict $R_{H}(Q)$ (see Fig. \ref{fig:figure1} (c)) with the dose at stake marked by a red data point. The origin of the coordinate system of this inset figure has been set to the CNP. \textbf{(i)} Conductivity $\Delta\sigma_{xx}(B)$ measured on sample A at the CNP fitted using the HLN formula.}
	\label{fig:figure3}
\end{figure*}
Figs. \ref{fig:figure3} (a) - (h) depict the normalized magnetoresistance of sample A measured as a function of magnetic field for selected $Q$. For the unirradiated sample, the magnetoresistance is parabolic at low fields. As $Q$ increases and as one approaches the maximum $R_{H}(Q)$, the magnetoresistance at low fields becomes negative, whereas at high fields it remains parabolic. After passing the maximum of $R_{H}(Q)$, the negative magnetoresistance at low fields persists, but in addition to that a sharp negative correction appears around zero field and is most pronounced around the CNP (see Figs. \ref{fig:figure3} (d)-(f)). For doses above the CNP, i.e., $R_{H}(Q)<0$, the negative magnetoresistance disappears. The sharp cusp around zero field becomes less pronounced with increasing dose until it disappears and the magnetoresistance at low fields becomes parabolic again.\\
A narrow regime of negative magnetoresistance at low magnetic fields is often considered as the hallmark of weak localization (WL) and arises in weakly disordered systems due to quantum interference of scattered electron waves when the electronic transport is in the \textit{diffusive regime}, i.e., on the metallic side of the Mott insulator transition (see below). WL has been observed in a large number of heavily doped semiconductors such as Si \cite{Roth:1963}, Ge \cite{Roth:1963,Woods:1964} or GaAs \cite{Woods:1964}. The dose dependence of the observed WL effect is intriguing. It should be noted that that in special cases WL has been proposed to originate from the bulk of a three-dimensional TI \cite{Garate:2012}. Garate et al. predict that the bulk of a 3D TI can show WL if the Fermi energy is close to the band edge, which would in fact agree with our observation of WL in a limited dose range inside region II. However, the calculations have been performed under numerous assumptions that are not fulfilled regarding our Bi$_2$Te$_3$ samples (band gap at the $\Gamma$-point, sample thinner than the phase coherence length). In all of the above cases, the observed negative magnetoresistance is associated to bulk carriers (see also Supplemental Material, Section C).\\
The negative correction to the magnetoresistance observed at low fields around the CNP is the characteristic signature for weak antilocalization (WAL). WAL has been by now observed in a large number of studies of TIs and is associated with the topologically protected states on the TI surface \cite{Brahlek:2014B,Chen:2011,Steinberg:2011,He:2011}. The magnetic field dependence of the WAL conductivity correction for the surface of a 3D TI is identical to the Hikami, Larkin, Nagaoka (HLN)-formula derived for strong strong spin-orbit coupling (SOC) \cite{Hikami:1980,Tkachov:2011}, i.e., $\Delta\sigma_{xx}=(\alpha e^2/\pi h) \left[ \Psi\left(1/2 + B_{\phi}/B\right) - \textnormal{ln} \left( B_{\phi}/B \right) \right]$ with $\Psi$ the digamma function, $B_{\phi}=\hbar/(4 e l_{\phi}^{2})$ and $l_{\phi}$ the dephasing length. The coefficient $\alpha$ is predicted to be $-1/2$ for a 2D transport channel in the presence of strong SOC \cite{Hikami:1980} or one TI surface \cite{Tkachov:2011}. Fig. \ref{fig:figure3} (i) plots $\Delta\sigma_{xx}=\Delta\sigma_{xx}(B)-\Delta\sigma_{xx}(B=0)$ measured at the CNP ($Q=18.7$ mC/cm$^2$) together with the fit using the HLN-formula yielding $\alpha=-(0.99\pm0.05)$ and $l_{\phi}=(370\pm5)$ nm. The calculated $l_{\phi}$ is comparable to the values usually obtained for mesoscopic TI samples \cite{Brahlek:2014B,Chen:2011,Steinberg:2011,He:2011} and much smaller than the sample thickness. An $\alpha\approx-1$ can be associated with either two decoupled TI surface states at the top and bottom surface of the sample or two 2D transport channels with SOC. Recently, several studies confirmed the formation of a topologically trivial 2D electron gas at the surface of TIs due to band bending, which can give rise to WAL as well \cite{Bianchi:2010,ViolBarbosa:2013,Lee:2014}. An unambiguous distinction between these two possible origins which can also coexist is thus difficult and makes the WAL effect less suited as an unambiguous fingerprint for the topologically protected surface states (for more details, see Supplemental Material, Section D).\\
The coexistence of electrons and holes around the CNP evidenced by the Hall effect can be understood in terms of the theory of completely compensated semiconductors (CCSs) \cite{Shklovskii:1972,Shklovskii:1984}. In a CCS, random spatial inhomogeneities in the distribution of acceptor- and donor-type dopants, i.e., here native and irradiation-induced defects, cause fluctuations in the charge distribution and the Coulomb potential associated with these charges. Due to the vanishing number of free charge carriers near complete compensation, these fluctuations are poorly screened and can locally bend the valence and conduction band edges. As illustrated in Fig. \ref{fig:figure4} (a), this ultimately results in the formation of hole and electron \textit{puddles}, if the valence or conduction band edge is bent above or below the Fermi energy, respectively. The carriers appearing in these regions will then prevent the band from being bent any further. This limits the amplitude and the range of the potential fluctuations which is characterized by the non-linear screening radius $R_g = E_{g}^{2}\kappa^{2}/N_{t}e^{4}$ with $E_{g}$ the bandgap, $\kappa$ the dielectric constant of the material, $N_t=N_A+N_D$ the total dopant concentration and $N_A$ ($N_D$) the acceptor- (donor-) dopant concentration \cite{Shklovskii:1972,Shklovskii:1984}. In the unirradiated $p$-type samples, the concentration of Bi$_{\textnormal{Te1}}$ antisite defects roughly equals the hole carrier concentration since these act as single acceptors \cite{Scanlon:2012}. Irradiation introduces donor-type defects, while the concentration of Bi$_{\textnormal{Te1}}$ antisite defects is not expected to change significantly during irradiation. With $\epsilon_{gap}=0.16$ eV \cite{Chen:2009}, $\kappa\approx100$ \cite{Goltsman:1985} and $N_{t}=2p_{0}$ one obtains $R_g=15$ and 1.4 $\mu$m for samples A and B at the CNP, respectively. The size of the hole and electron puddles (shaded regions in Fig. \ref{fig:figure4} (a)) can be estimated by $R_{h,e}=a_{h,e}/{(N_{t}a_{h,e}^3)}^{1/9}$ \cite{Shklovskii:1972,Shklovskii:1984} with $a_{h,e}=\kappa m_{0}a_{0}/m^{\ast}_{h,e}$ the effective Bohr radius of the holes/electrons, $a_0=0.529$ \AA$~$the Bohr radius of the hydrogen atom and $m^{\ast}_{h,e}$ the effective mass of the carriers ($m^{\ast}_{h}=0.08m_{0}$ \cite{Koehler:1976}, $m^{\ast}_{e}=0.06m_{0}$ \cite{Koehler:1976a}).  One obtains similar sizes for hole and electron puddles which amount to $R_{h,e}\approx31$ and 24 nm for samples A and B, respectively. $R_{h,e}$ is magnitudes smaller than $R_g$, i.e., a large number of puddles always contributes to the screening of the potential fluctuations \cite{Shklovskii:1972,Shklovskii:1984}. The average distance between defects is only about $\sqrt[3]{1/N_{t}}=5$ and 2 nm for samples A and B, i.e., each puddle hosts many defects.\\
For both samples, the range $R_g$ of the potential fluctuations is extremely large. It should be stressed that these fluctuations are \textit{always} present in compensated samples independent of the method of compensation, i.e., irradiation or chemical doping \cite{Borgwardt:2015}. Their range $R_g$ and the size of the puddles is completely determined by the number of defects in the uncompensated state, i.e., here $N_A$. For a higher initial defect concentration (sample B) one obtains fluctuations with a shorter range $R_g$ and a smaller puddle sizes at the CNP. It should be noted that the calculated values for $R_{h,e}$ agree well with what has been found in STM measurements on chemically doped TIs \cite{Beidenkopf:2011}.\\
\begin{figure*}
	\centering
		\includegraphics[width=0.99\textwidth]{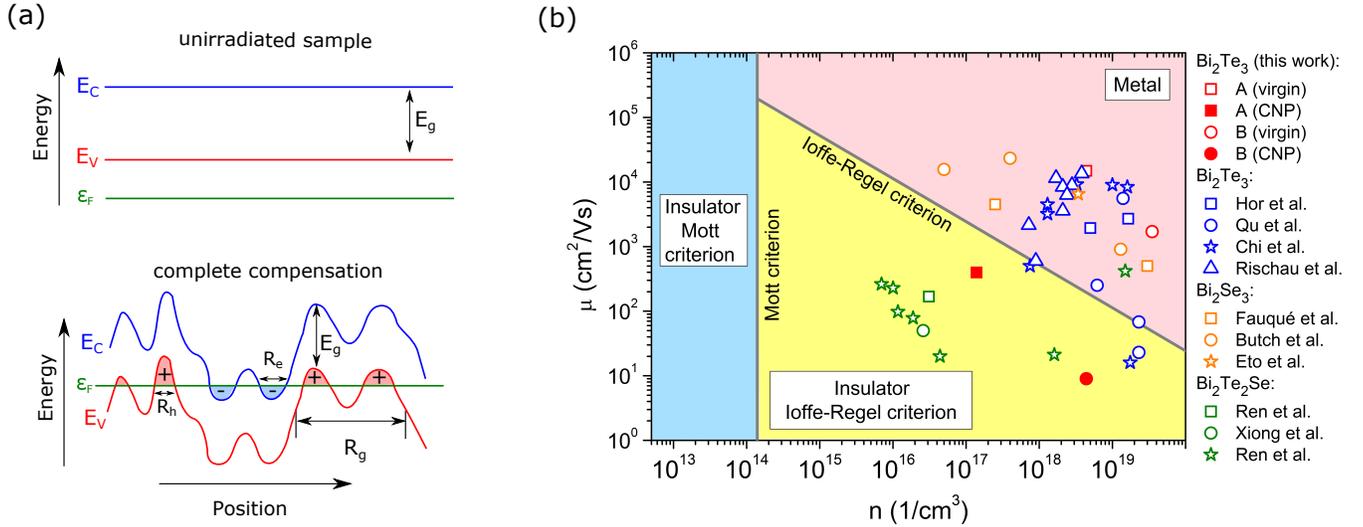}
	\caption{\textbf{(a)} Formation of charge puddles in a CCS with band gap $E_g$ and valence and conduction band edge $E_v$ and $E_c$ (figure adapted with permission from \cite{Skinner:2012}. Copyrighted by the American Physical Society). In the unirradiated sample the conduction is $p$-type and the Fermi energy lies in the valence band. With increasing compensation the Fermi energy moves inside the gap and electron and hole puddles form in places where the valence and conduction band (meandering lines) are bent above or below the Fermi energy $\epsilon_F$, respectively. \textbf{(b)} Illustration of the Mott and the Ioffe-Regel criterion (see also Ref. \cite{Brahlek:2014B}) for Bi$_2$Te$_3$ samples irradiated at low (this work) and room temperature \cite{Rischau:2013}. Data obtained on as grown Bi$_2$Te$_3$, Bi$_2$Se$_3$ and Bi$_2$Te$_2$Se bulk samples in Refs. \cite{Qu:2010,Hor:2010,Chi:2013,Fauque:2013,Butch:2010,Eto:2010,Ren:2010,Xiong:2012,Ren:2012} is shown for comparison.}
	\label{fig:figure4}
\end{figure*}
The question arising at this point, is whether or not irradiation can be used as an effective new way to create TI samples with an insulating bulk? Inspired by the discussion in a recent work by Brahlek et al. \cite{Brahlek:2014B},  Fig. \ref{fig:figure4} (b) plots the carrier mobility vs. carrier density of the irradiated samples as well as data obtained on TI \textit{bulk} samples in Refs. \cite{Qu:2010,Hor:2010,Chi:2013,Fauque:2013,Butch:2010,Eto:2010,Ren:2010,Xiong:2012,Ren:2012}. Fig. \ref{fig:figure4} (b) further visualizes when a metal-insulator transition would be expected according the Mott \cite{Mott:1956} or the Ioffe-Regel \cite{Ioffe:1960} criterion. Based on a dielectric screening approach, Mott predicted that a material will undergo a transition from the metallic to the insulating state (\textit{Mott insulator}) for dopant densities $N<N_c$. This \textit{Mott criterion} predicts the critical dopant concentration $N_c$ as $a_{h,e}N_{c}^{1/3}\approx0.25$ and has been found to apply to a large number of doped semiconductors \cite{Edwards:1978}. For Bi$_2$Te$_3$, one can estimate $N_{c}\approx1.4\times10^{14}$ cm$^{-3}$, which is quite low compared to other semiconductors as for example Si ($N_{c}\approx2\times10^{18}$ cm$^{-3}$). Ioffe and Regel predicted that a metal-insulator transition will occur if a decrease of the carrier density is accompanied by an increase of disorder \cite{Ioffe:1960}. According to this \textit{Ioffe-Regel criterion} a transition from the metallic ($k_{F}l\gg1$) to the insulating ($k_{F}l \ll 1$) state occurs at $k_{F}l\approx1$ with $k_F$ the Fermi wave vector and $l$ the mean free path. For strong disorder, diffusive electron motion is no longer possible and at $T=0$, the carriers are truly localized (\textit{Anderson insulator}). Simplifying the six-valley Fermi surface of Bi$_2$Te$_3$ \cite{Koehler:1976} by assuming spherical hole pockets with $k_F=(3\pi^{2}n/6)^{1/3}$ and $l=\frac{\hbar}{e}k_{F}\mu$, one can express $k_{F}l=1$ as $\mu=\frac{e}{\hbar}(\pi^{2}n/2)^{-2/3}$. It should be noted that the Ioffe-Regel criterion assumes uniform systems and is not strictly valid in the case of strongly compensated TIs. With increasing degree of compensation and the formation of puddles, classical percolation of carriers and metallic conductivity are lost \cite{Shklovskii:1972,Shklovskii:1984}. As can be seen in Fig. \ref{fig:figure4} (b), the irradiated samples and all bulk Bi$_2$Te$_3$, Bi$_2$Se$_3$ or Bi$_2$Te$_2$Se samples that were studied so far, are far from being insulators in the Mott sense. Irradiation results in a simultaneous decrease of carrier density and mobility and at the CNP, the irradiated samples are more insulating than most of the as-grown Bi$_2$Te$_3$ and Bi$_2$Se$_3$ samples, however, more metallic than Bi$_2$Te$_2$Se \cite{Ren:2010,Xiong:2012,Ren:2012}. Furthermore, the magnetoresistance measurements for doses around the CNP display a signature of weak antilocalization very similar to what has been observed on thinned-down, electrically gated or chemically doped TI samples and what is considered to be the hallmark of the topologically protected surface states.\\
In summary, we showed that compensation of the intrinsic charge carriers in Bi$_2$Te$_3$ leads to the confinement of bulk carriers into charge puddles. These puddles will always be present in compensated samples independent of the method of compensation and need to be considered to understand transport in compensated TIs.\\
We thank the \textit{SIRIUS} team for technical support during the irradiations and A. Hruban, A. Wolos and M. Kaminska for supplying one of the samples.

\end{document}